\documentclass[Physsubmission, Phys]{SciPost}
\hypersetup{
    colorlinks,
    linkcolor={red!50!black},
    citecolor={blue!50!black},
    urlcolor={blue!80!black}
}

\usepackage[bitstream-charter]{mathdesign}
\DeclareSymbolFont{usualmathcal}{OMS}{cmsy}{m}{n}
\DeclareSymbolFontAlphabet{\mathcal}{usualmathcal}

\begin{document}

\begin{center}{\Large \textbf{
What do DVCS data tell us about TCS observables?\\
}}\end{center}

% TODO: write the author list here. Use initials + surname format.
% Separate subsequent authors by a comma, omit comma at the end of the list.
% Mark the corresponding author with a superscript *.
\begin{center}
    O.~Grocholski\textsuperscript{1,2}
    , H.~Moutarde\textsuperscript{3} 
    , B. Pire\textsuperscript{4} 
    , P. Sznajder\textsuperscript{1}
    , J. Wagner\textsuperscript{1$\star$}
\end{center}

\begin{center}
{\bf 1} National Centre for Nuclear Research (NCBJ), Pasteura 7, 02-093 Warsaw, Poland
\\
{\bf 2} Institute of Theoretical Physics, Faculty of Physics, University of Warsaw, Pasteura 5, 02-093 Warsaw, Poland
\\
{\bf 3} IRFU, CEA, Université Paris-Saclay, 91191 Gif-sur-Yvette, France
\\
{\bf 4} CPHT, CNRS, École Polytechnique, I. P. Paris, 91128 Palaiseau, France
\\
* jakub.wagner@ncbj.gov.pl
\end{center}

\begin{center}
%\today
\end{center}

% For convenience during refereeing (optional),
% you can turn on line numbers by uncommenting the next line:
%\linenumbers

\definecolor{palegray}{gray}{0.95}
\begin{center}
\colorbox{palegray}{
  \begin{tabular}{rr}
  \begin{minipage}{0.1\textwidth}
    \includegraphics[width=22mm]{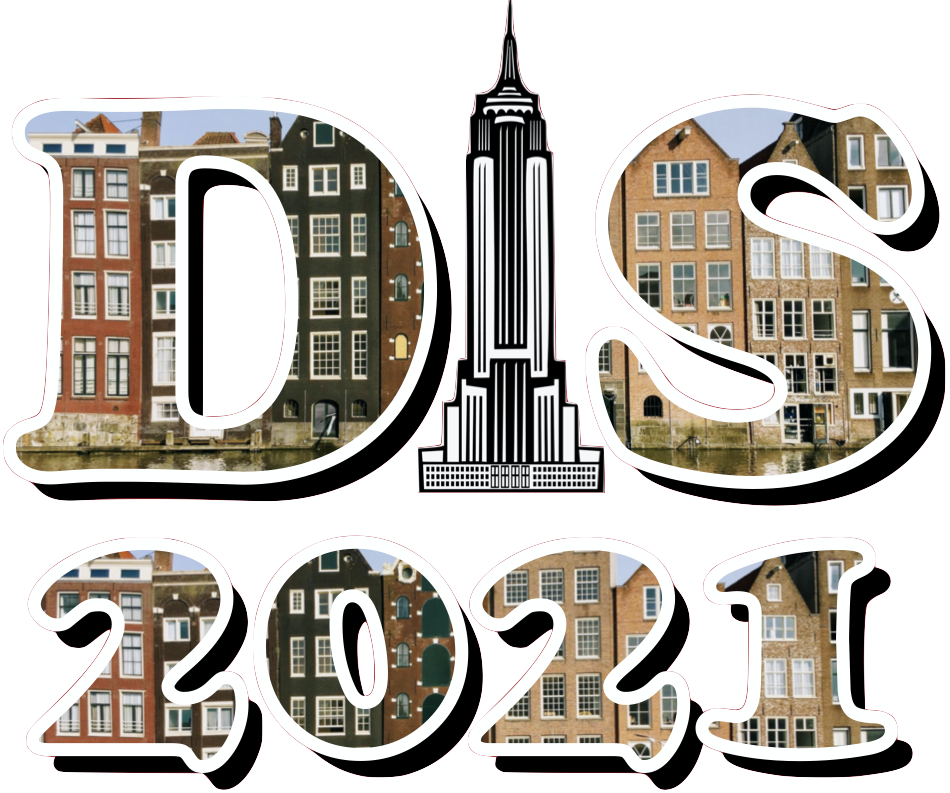}
  \end{minipage}
  &
  \begin{minipage}{0.75\textwidth}
    \begin{center}
    {\it Proceedings for the XXVIII International Workshop\\ on Deep-Inelastic Scattering and
Related Subjects,}\\
    {\it Stony Brook University, New York, USA, 12-16 April 2021} \\
    \doi{10.21468/SciPostPhysProc.?}\\
    \end{center}
  \end{minipage}
\end{tabular}
}
\end{center}

\section*{Abstract}
{\bf
% TODO: write your abstract here.
%Using a recent extraction of 
Deeply virtual Compton scattering (DVCS) and  
%Compton form factors, done within the PARTONS framework, we derive 
timelike Compton scattering (TCS) leading twist amplitudes 
%and calculate TCS observables only assuming leading-twist dominance. In the framework of collinear QCD factorization, the leading-twist scattering amplitudes for DVCS and TCS 
are intimately related thanks to their analytic properties as a function of $Q^2$.
%of leading and next-to-leading order amplitudes. 
We exploit this  feature to use  Compton form factors previously extracted from available DVCS data and derive data-driven predictions for TCS observables to be measured in near future experiments.
Our results quantitatively illustrate the complementarity of DVCS and TCS experiments.
%Artificial neural network techniques are used for an essential reduction of model dependency, allowing for stringent tests of the universality of leading-twist description of DVCS and TCS amplitudes in terms of Generalized Parton Distributions (GPDs). Moreover, this study helps to understand quantitatively the complementarity of DVCS and TCS measurements, which is crucial e.g. to perform the nucleon tomography. {\color{red} This needs to be changed, so far it is copy-paste from the paper.}
}

%\vspace{10pt}
%\noindent\rule{\textwidth}{1pt}
%\tableofcontents\thispagestyle{fancy}
%\noindent\rule{\textwidth}{1pt}
%\vspace{10pt}

\section{Introduction}
\label{sec:intro}
% TODO: write your article here.
Deeply virtual Compton scattering (DVCS) 
\begin{equation}
\gamma^*(q) N(p_1) \to \gamma(q') N'(p_2) \,~~~\text{at large}~~ Q^2=-q^2 \, ,
\end{equation}
and  
timelike Compton scattering (TCS) 
\begin{equation}
\gamma(q) N(p_1) \to \gamma^*(q') N'(p_2) \,~~~\text{at large}~~ Q'^2=+q'^2 \,,
\end{equation}
are the two facets of the simplest application of QCD collinear factorization techniques to exclusive hard amplitudes in the near-forward (small $t=(p_2-p_1)^2$) generalized Bjorken regime. While many experimental data have already been reported for DVCS \cite{dHose:2016mda,Kumericki:2016ehc}, TCS experimental studies did only very recently report their first results \cite{Chatagnon:2020nvc}. The aim of our study~\cite{Grocholski:2019pqj} is to quantify the complementarity of DVCS and TCS data to extract information on generalized parton distributions (GPDs), in as much as possible a model-independent way, helped by artificial neutral network techniques.

\section{Relation between DVCS and TCS amplitudes}
\label{sec:DVCS_vs_TCS}
The straightforward intimate connection between DVCS and TCS amplitudes was noticed in the early papers \cite{Mueller:1998fv, Berger:2001xd} and made explicit at the next to leading order (NLO) in $\alpha_s$ in \cite{Pire:2011st, Moutarde:2013qs}.
After a decomposition of scattering amplitudes in products of Dirac structures and Compton form factors (CFF), factorization theorems allow to express these CFFs in terms of perturbatively calculable coefficient functions $T^i$ and GPDs $F^i$, where $i$ denotes the various parton types:
\begin{eqnarray}
{\cal F}(\xi, t,{\cal Q}^2) =
\int_{-1}^1\!\! dx\!
\sum_{i=u,d,\ldots, g}\!\! {T^i(x,\xi,{\cal Q}^2)} {F^i(x,\xi,t)} \,.
\label{eq:factorizedamplitude}
\end{eqnarray}
The simple spacelike-to-timelike relations derived in Ref.~\cite{Muller:2012yq} allow us to express the NLO timelike coefficients by the spacelike ones as:
\begin{eqnarray}
{^T}T^i &\stackrel{\rm NLO}{=}& \pm {^S}T^{i\,*} \mp i\pi \frac{\alpha_s(\mu_{R}^2)}{2\pi} {^S}C_{\mathrm{coll}}^{i\,*} \,, 
\label{fullrelation}
\end{eqnarray}
where left superscripts $S$ and $T$ respectively denote spacelike and timelike quantities, and upper (lower) sign is for (anti-)symmetric coefficient functions in $\xi$;  ${^SC_{\mathrm{coll}}^i}$ can be found in Ref. \cite{Muller:2012yq}. For (anti-)symmetric CFFs $\cal H$ ($\widetilde{\cal H}$) this gives:
\begin{eqnarray}
{^T{\cal H}}  \stackrel{\rm NLO}{=} + ^S{\cal H}^\ast - i\pi\, {\cal Q}^2\frac{\partial}{\partial {\cal Q}^2} {^S{\cal H}}^\ast \,~~~,~~~
{^T\widetilde{\cal H}}  &\stackrel{\rm NLO}{=}&  -^S\widetilde{\cal H}^\ast + i\pi\, {\cal Q}^2\frac{\partial}{\partial {\cal Q}^2}  {^S\widetilde{\cal H}^\ast}\,,
\label{eq:cffHt2cffTHt}
\end{eqnarray}
and similar relations  for (anti-)symmetric CFFs $\cal E$ ($\widetilde{\cal E}$). Let us remind the reader that the LO CFF depends only on quark GPDs while gluon GPDs contributions enter NLO CFFs. 
\section{Data-driven predictions for TCS observables}
\label{sec:TCS_observables}

\begin{figure}[ht]
\includegraphics[width=0.49\linewidth]{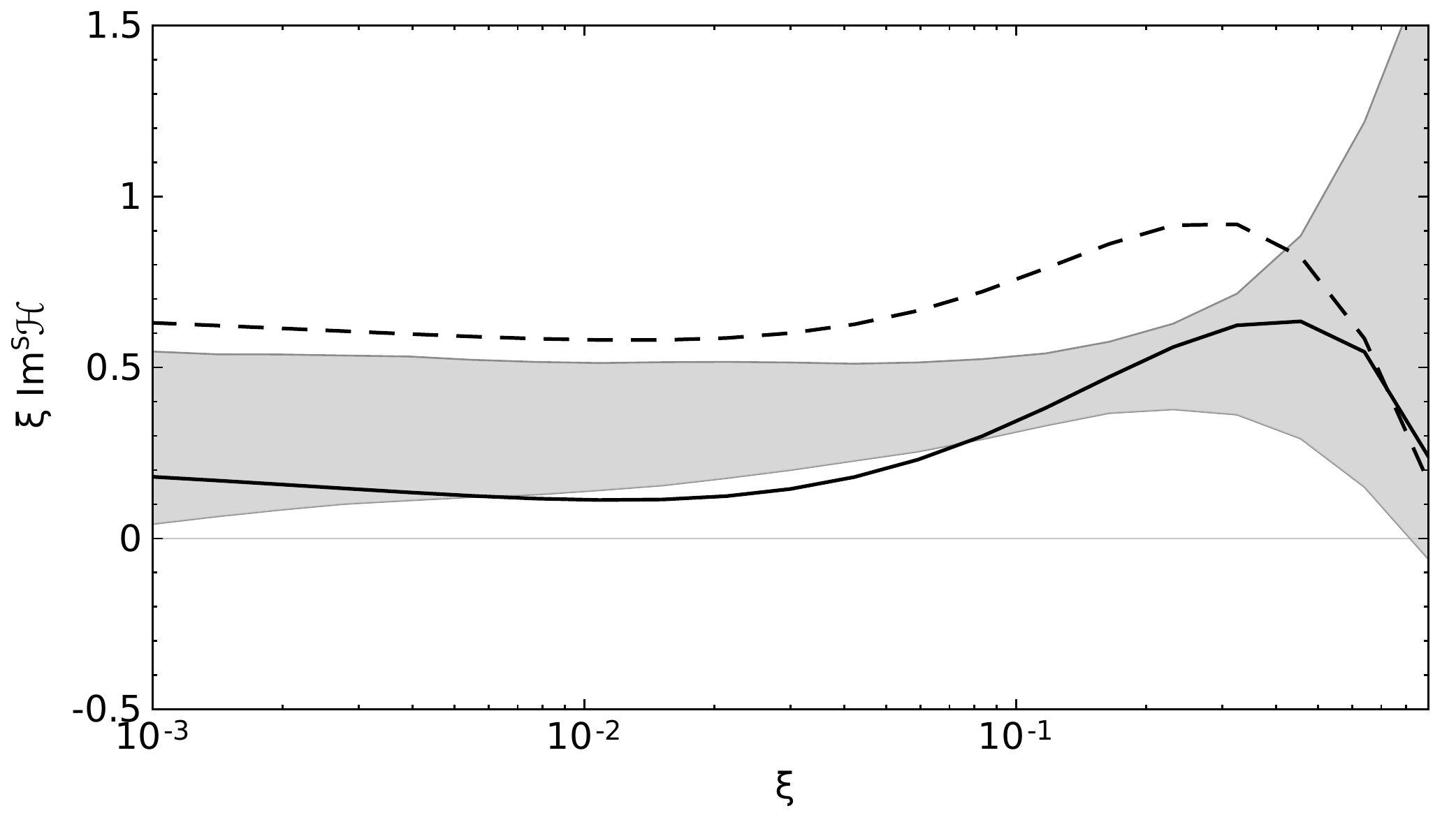}
\includegraphics[width=0.49\linewidth]{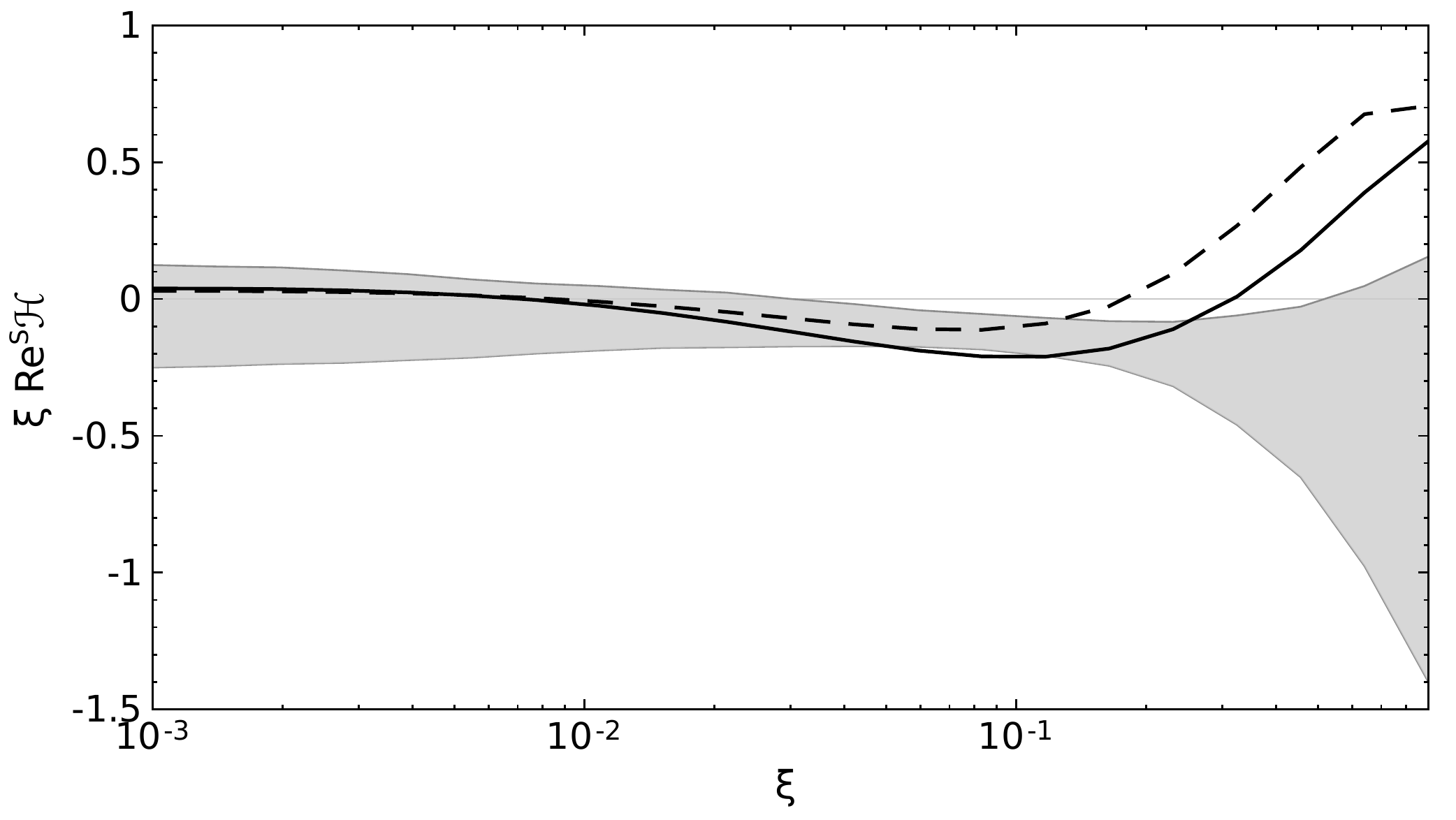}
\caption{Imaginary (left panel) and real (right panel) part of DVCS CFF $\xi ^S\mathcal{H}(\xi)$ for $Q^2 = 2$ GeV$^2$ and $t=-0.3$ GeV$^2$ as a function of $\xi$. The shaded gray bands correspond to the global fit of DVCS data presented in \cite{Moutarde:2019tqa} and they show 68\% confidence level for the uncertainties of presented quantities. The dashed (solid) lines correspond to the GK GPD model \cite{Goloskokov:2005sd, Goloskokov:2007nt, Goloskokov:2009ia} evaluated with LO (NLO) DVCS coefficient functions.}
\label{fig:DVCS_CFFs}
\end{figure}

%%%%%%%%%%%%%%%%%%%%%%%%
\begin{figure}[!ht]
\begin{center}
\includegraphics[width=0.49\linewidth]{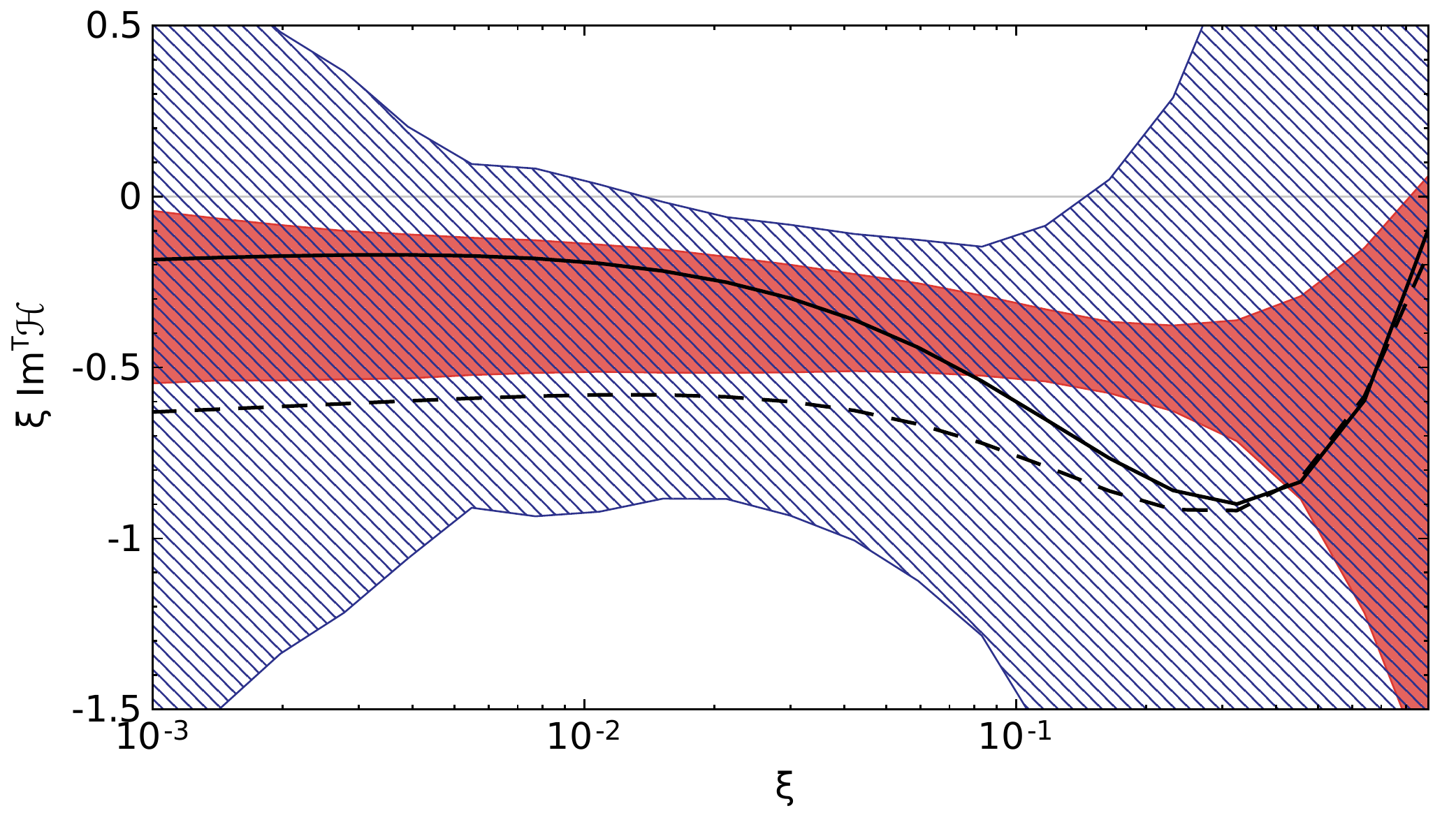}
\includegraphics[width=0.49\linewidth]{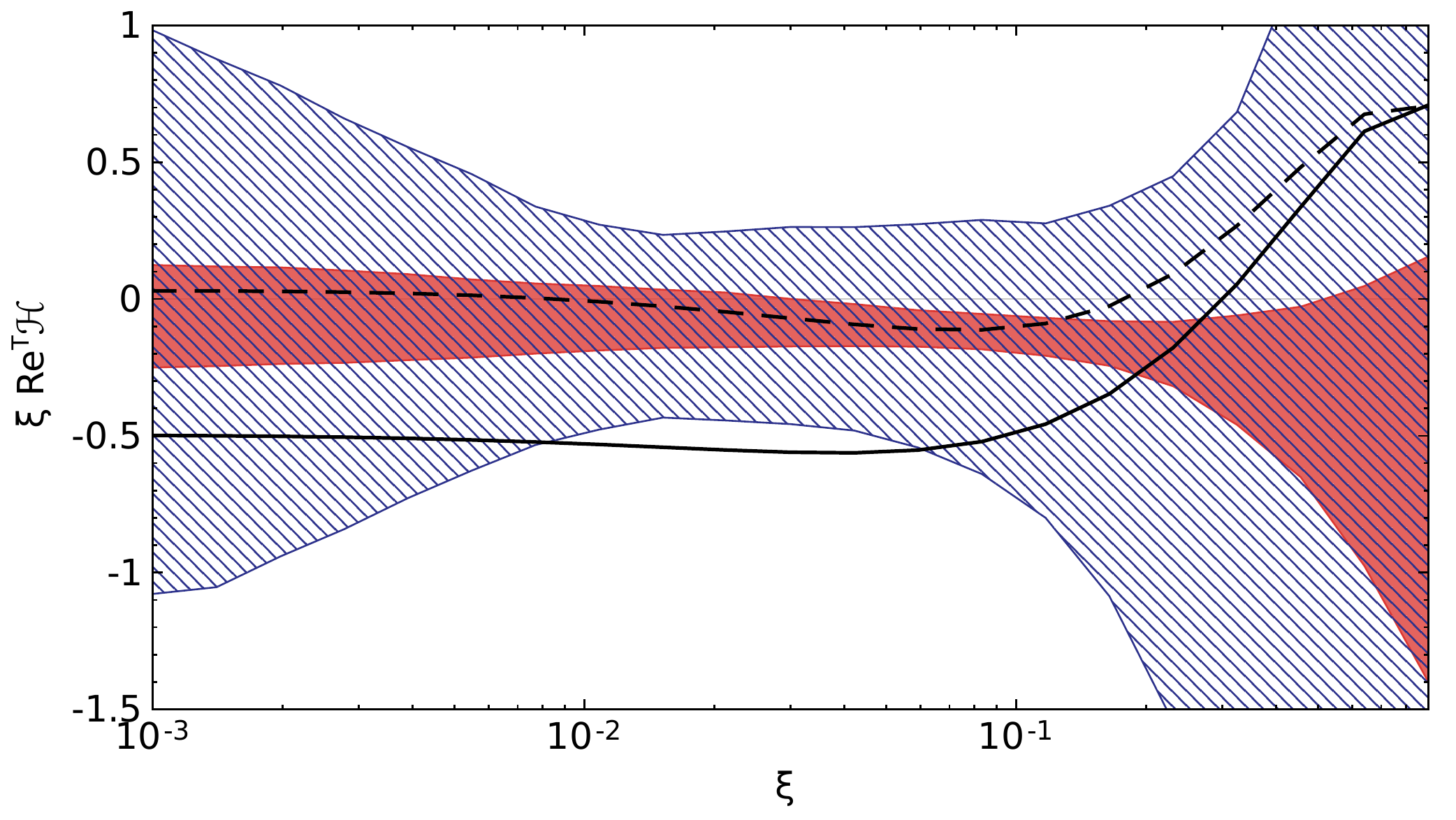}
\caption{Imaginary (left) and real (right) part of TCS CFF $\xi ^T\mathcal{H}(\xi)$ for $Q^2 = 2$ GeV$^2$ and $t=-0.3$ GeV$^2$ as a function of $\xi$. The shaded red (dashed blue) bands correspond to the data-driven predictions coming from the global fit of DVCS data presented in \cite{Moutarde:2019tqa} and they are evaluated using LO (NLO) spacelike-to-timelike relations. The bands show 68\% confidence level for the uncertainties of presented quantities. The dashed (solid) lines correspond to the GK GPD model \cite{Goloskokov:2005sd, Goloskokov:2007nt, Goloskokov:2009ia} evaluated with LO (NLO) TCS coefficient functions.}
\label{fig:TCS_CFFs}
\end{center}
\end{figure}
%%%%%%%%%%%%%%%%%%%%%%%%
Our basic tool for deriving data-driven predictions for TCS is to use the artificial neural network technique  employed in \cite{Moutarde:2019tqa} to determine the spacelike CFFs from a global analysis of almost all DVCS measurements off a proton target. This technique is known to lead to an essential reduction of model dependency. In our analysis the replica method was used to propagate experimental uncertainties to those of extracted quantities.
Let us now present some illustrative results (see \cite{Grocholski:2019pqj} for more plots). In Fig. \ref{fig:DVCS_CFFs}  we  show the extracted DVCS CFF $^S\cal{H}$ (shaded gray band) as a function of $\xi$ for exemplary kinematics of $Q^2 =2$~GeV$^2$, $t = -0.3$ GeV$^2$. For comparison, we also present a model prediction based on the Goloskokov-Kroll (GK) parametrization of GPDs \cite{Goloskokov:2005sd, Goloskokov:2007nt, Goloskokov:2009ia}, obtained with LO (dashed line) and NLO (solid line) coefficient functions.

%In these plots, the lines correspond to model-dependent predictions coming from the GPD model \cite{Goloskokov:2005sd, Goloskokov:2007nt, Goloskokov:2009ia}, which is consistent with a lO analysis of DVCS data. The bands correspond to our unbiased way to extract CFFs from DVCS data.  

\begin{figure}[ht]
\begin{center}
\includegraphics[width=0.5\linewidth]{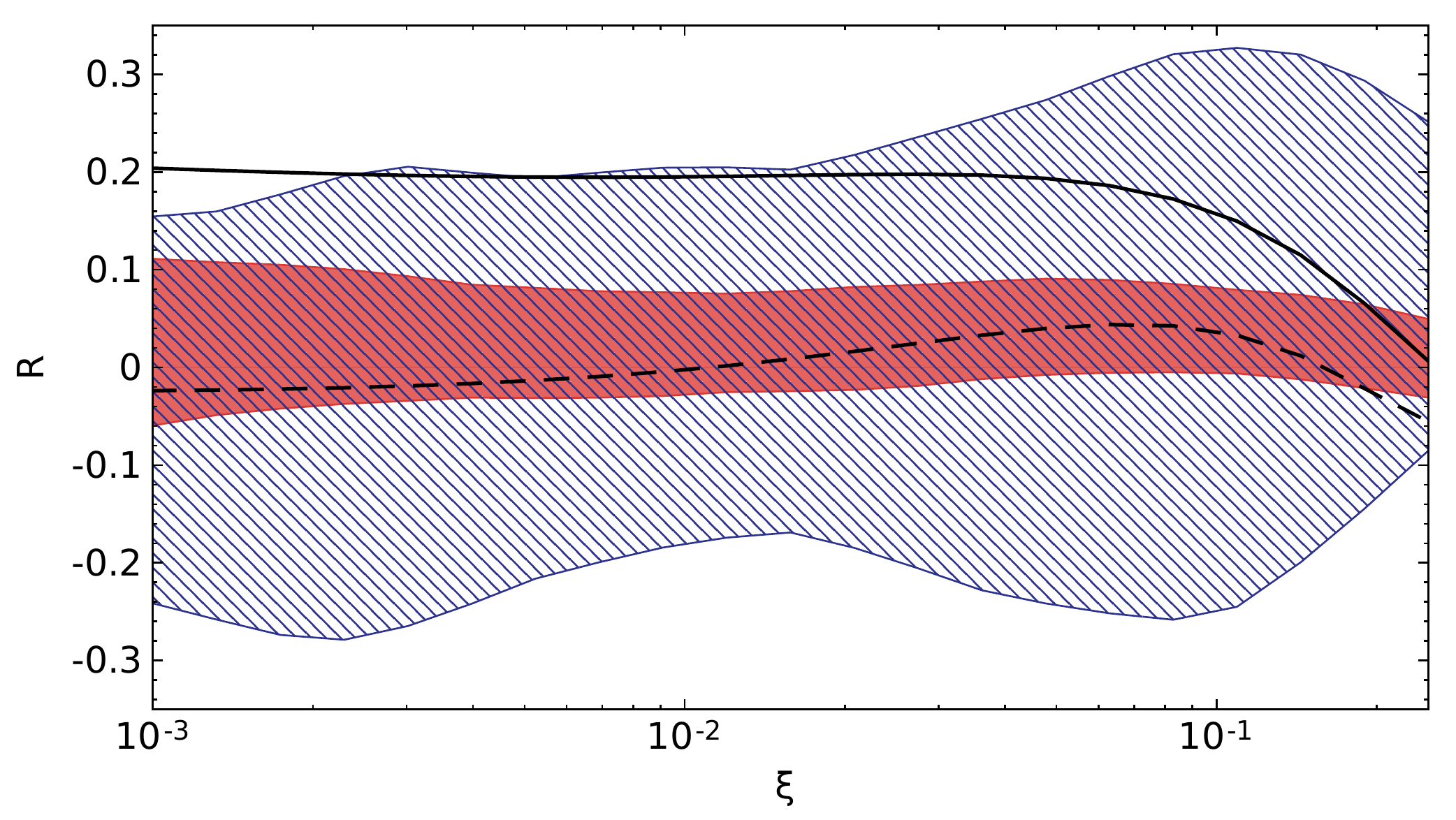}
\caption{Ratio $R$ defined in \cite{Berger:2001xd} evaluated with LO and NLO spacelike-to-timelike relations for $Q'^2 = 4$ GeV$^2$, $t=-0.35$~GeV$^2$ as a function of $\xi$. For  further description see the caption of Fig. \ref{fig:TCS_CFFs}.
}
\label{fig:R}
\end{center}
\end{figure}

Fig.\ref{fig:TCS_CFFs} shows the $\xi$-dependence of the dominant TCS, CFF $\xi ^T\mathcal{H}(\xi)$ for $Q^2 = 2$ GeV$^2$ and $t=-0.3$ GeV$^2$ using both LO and NLO relations between the DVCS and TCS amplitudes. The bigger uncertainty attached to the NLO case is representative of the very bad knowledge of the $Q^2$ dependence of DVCS amplitudes,  reflecting the sparsity and limited range in $Q^2$ of the used data. Turning the argument around, this indicates that a moderately precise TCS measurement would help much in quantifying this $Q^2$ dependence which is of utmost importance to fully understand the QCD dynamics of these processes.
\begin{figure}[ht]
\begin{center}
\includegraphics[width=0.5\linewidth]{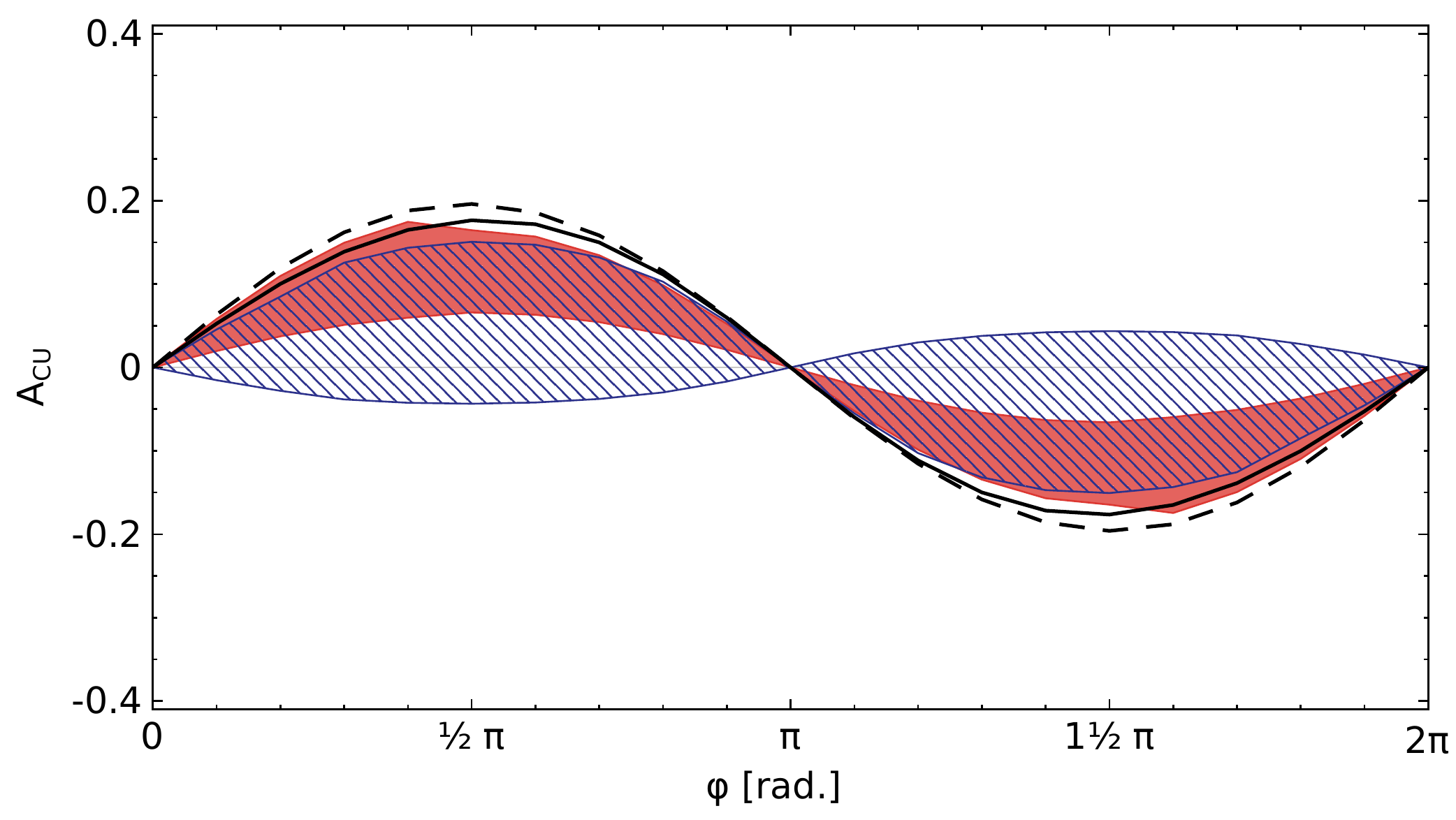}
\end{center}
\caption{Circular asymmetry $A_{CU}$ evaluated with LO and NLO spacelike-to-timelike relations for $Q'^2 = 4$ GeV$^2$, $t=-0.1$~GeV$^2$ and $E_\gamma =10$ GeV as a function of $\phi$.  The cross sections used to evaluate the asymmetry are integrated over $\theta\in (\pi/4,3\pi/4)$. For  further description see the caption of Fig. \ref{fig:TCS_CFFs}.
}
\label{fig:ACU_phi}
\end{figure}

In Fig.\ref{fig:R}, we show our data-driven predictions for the ratio $R$ defined in \cite{Berger:2001xd}, which is particularly interesting since it projects out the interference term between the BH and TCS amplitudes, that is linear in CFFs, and which has a special sensitivity to the real part of CFF $^T\cal{H}$.

Fig.\ref{fig:ACU_phi} displays our predictions for the azimuthal angle dependence of the circular photon polarization asymmetry
\begin{equation}
A_{CU}(\phi) = \frac{\sigma(\nu = +1)-\sigma(\nu = -1)}{\sigma(\nu = +1)+\sigma(\nu = -1)} \,.    
\end{equation}
which singles out specific elements of the interference contribution to the cross section. The denominator of this asymmetry is dominated by the square of BH amplitude, which is almost flat in $\phi$.

\section{Conclusion}
Our data-driven study of TCS has demonstrated how much it is important to use  both DVCS and TCS data to access in a sensible way the GPDs of the nucleon. It also showed in a quantitative way why any extraction of GPDs  based on a leading order analysis of experimental data is very incomplete and  model-dependent.  In particular, the analytical (in $Q^2$) relation between NLO coefficient functions of DVCS and TCS results in a quite unique way to access the $Q^2$ dependence of the GPDs. Other reactions should be analyzed in the same way. In particular the cases of double DVCS \cite{Guidal:2002kt,Belitsky:2002tf} and of large invariant mass diphoton photoproduction \cite{Pedrak:2017cpp}, which are also pure QED processes at leading order, are quite interesting since there both a timelike and a spacelike large invariant set the factorization scale. 

\section*{Acknowledgements}
This project was supported by the European Union's Horizon 2020 research and innovation programme under grant agreement No 824093 and by the Grant No. 2017/26/M/ST2/01074 of the National Science Centre, Poland. The project is co-financed by the Polish National Agency for Academic Exchange and by the COPIN-IN2P3 Agreement.

%\bibliography{SciPost_Wagner.bib}

%\nolinenumbers

\end{document}